# Imaging Josephson Vortices on Curved Junctions

Yuita Fujisawa[1], Anjana Krishnadas[1], Tomonori Nakamura[1], Chia-Hsiu Hsu[1,2], Barnaby Smith[1], Markel Pardo-Almanza[1], Hoshu Hiyane[1], Guoqing Chang[2], Yuki Nagai[3,4], Tadashi Machida[5], Yoshinori Okada[1]

[1]*Quantum Materials Science Unit, Okinawa Institute of Science and Technology (OIST), Okinawa 904-0495, Japan.*
[2]*Division of Physics and Applied Physics, School of Physical and Mathematical Sciences, Nanyang Technological University, 637371 Singapore*
[3]*Information Technology Center, The University of Tokyo, Chiba 277-0882, Japan*
[4]*CCSE, Japan Atomic Energy Agency, 178-4-4, Wakashiba, Kashiwa, Chiba, 277-0871, Japan*
[5]*RIKEN Center for Emergent Matter Science, Saitama, 351-0198, Japan*

Understanding the nature of vortices in type-II superconductors is crucial for comprehending exotic superconductors and advancing the application of superconducting materials in future electronic devices. This study uses spectroscopic scanning tunneling microscopy to visualize Josephson vortices along crystalline domain boundaries in the superconducting spinel oxide $LiTi_2O_4$ (LTO). Our experimental results reveal that the local curvature of the Josephson junction dictates the positioning of Josephson vortices. Self-consistent solutions of the Bogoliubov-de Gennes and gap equations theoretically corroborate this observation. In addition to enhancing our understanding of the physics of Josephson vortex formation, this study offers potential guidelines for developing vortex-based superconducting devices.

Applying an external magnetic field ($B$) leads to a mixed state where Cooper pair breaking occurs partially, allowing the penetration of magnetic flux in type II superconductors. Since the motion of vortices causes non-zero resistance when electric current flows, understanding the pinning mechanism not to move the vortex is thus one of the most critical challenges for practical applications [1,2,3]. On the other hand, the manipulation of the vortex positions can be essential for next-generation efficient computations [4,5,6,7]. Developing a novel principle or guideline for controlling vortices is desired for broad perspectives to advance superconducting electronic applications.

In a conventional Abrikosov vortex (Av), the primary pair-breaking is spatially confined to minimize the loss of superconducting condensation energy in association with accommodating a magnetic flux [see **Fig. 1(a)**]. The characteristic length scale of an Av is given by the coherence length ($\xi$) of the underlying superconducting order parameter. However, this scenario changes qualitatively when a Josephson junction (JJ) is introduced. A JJ comprises two adjacent superconducting regions separated by a non-superconducting barrier, allowing for a mismatch of phase of the order parameter ($\theta$) between separated superconducting regions. While a smooth phase rotation around the core is possible within a JJ when the junction width ($d$) is comparable to the coherence length ($\xi$) [see **Fig. 1(b)**], the impact of phase mismatch results in none trivial vortex formation for $d < \xi$ case [see **Fig. 1(c)**] [8,9,10]. At a qualitative level, the formation of an elongated Josephson vortex (JV) along the junction (JJ) is anticipated, as this minimizes the loss of superconducting condensation energy by selectively suppressing the weaker superconducting regions. Moreover, unlike the near-zero-bias conductance (ZBC) anomaly commonly observed in Andreev bound states (ABS) [11,12,13], the potential presence of bound state energy near the gap edge is also expected for a Josephson vortex (JV). This is because the binding energy of the quasiparticles excited within the vortex is approximately determined by the spatial extent of the vortex. Beyond these qualitative pictures, however, a quantitative understanding of the vortex adjacent to a JJ remains non-trivial. Given the intrinsic allowance for phase slips across the JJ, the total energy gain and loss due to the spatial evolution of $\theta(r)$ must be determined self-consistently, in conjunction with introducing a magnetic flux. Recently, the exotic nature of JVs was experimentally visualized on superconducting atomic sheets, where straight atomic steps acted as the JJ [14]. Self-consistent solutions of the Bogoliubov-de Gennes and gap equations theoretically supported the experimental observation. An intriguing remaining question concerns the behavior of JVs in curved junctions [see **Fig. 1(d)**], as phase variations around curved JJs may induce non-trivial effects on JV formation. However, this effect remains experimentally unexplored.

The material we shed light on in this study is a spinel oxide superconductor LiTi$_2$O$_4$ (LTO), whose structure is shown in **Fig. 1(e)**. The critical temperature ($T_c$) is 12 K, and the critical fields $B_{c1}$ and $B_{c2}$ are 0.03 and 11 T, respectively [15]. The previous scanning tunneling microscopy (STM) study on the LTO(111) epitaxial films demonstrated an atomically flat surface and an atomically sharp domain boundary (satisfying d < $\xi$ ) across which the crystalline axis is rotated by 180 degrees with respect to the surface normal (111) direction [16,17,18]. Here, focusing on inherent curved JJ (c-JJ) in this system [**Fig. 1(f)**], we visualized the individual vortexes on c-JJ [**Fig. 1(g)**]. Notably, we demonstrate the significant role of the curvature of JJ in controlling the position of the vortex, which is supported by a self-consistent solution of the Bogoliubov-de Gennes and gap equations.

In this study, the magnetic field-driven in-gap states are carefully investigated to visualize the JVs. **Figure 2(a)** shows the spatially averaged spectra at $B$ = 0, 1, and 1.5 T, obtained from the same field of view as **Fig. 1(f)**. The conductance near the gap edge is enhanced with increasing $B$ (pink and blue curves). Focusing on two energies within the gap [0 and +1.5 mV, see vertical lines in **Fig. 2(a)**], the series of the normalized conductance [$g$(0 mV)/$g$(+5 mV)] and [$g$(+1.5 mV)/$g$(+5 mV)] maps are shown in **Fig. 2(b,c,d)** and **Fig. 2(e,f,g)**, respectively. At $B$ = 0 T, the electronic state is essentially homogeneous at 0 mV and +1.5 mV [**Fig. 2(b, e)**]. On the other hand, prominent $B$-driven contrast is seen at $B$ = 1 and 1.5 T [**Fig. 2(c,d,f,g)**]. At 0 mV [**Fig. 2(c,d)**], the AVs are visualized, as we reported previously [17]. Intriguingly, additional contrast is observed near the c-JJ with the elongated shapes at +1.5 mV [**Fig. 2(f,g)**].

Three isolated contrasts in the $g$(+1.5 mV)/$g$(+5 mV) map at $B$ = 1 and 1.5 T are highlighted [denoted as JV #1-3 in **Fig. 2(f,g)**]. At the center of JV #1-3, the point spectra taken are also taken [**Fig. 2(h-j)**]. For comparison, the spectra taken at $B$ = 0 T at the same location (black dashed curves) and the spatially averaged spectrum at $B$ = 0 T (gray curves, same as **Fig. 2(a)**) are also overlaid. The identical shape of the two spectra at $B$ = 0 T (gray and black dashed curves) confirms that the conductance change under the magnetic field is unrelated to chemical segregation or disordering. At a finite $B$, the spectrum taken on JV #1-3 [**Fig. 3h-j**] show an enhanced conductance of approximately $\pm1.5$ mV, which is close to the gap edge energy. It has been theoretically known that the appearance of the bound state near the gap edge for vortex on JJ [12,13]. The validity of distinct bound state energy between AV and JV is also explained based on theoretical simulation in the latter part of this report.

The most important observation in this report is the location of JVs related to the local curvature of the c-JJs. As shown in **Fig. 2(f)**, JV #1 and #2 are located outside the junction's curvature. On the other hand, JV #3 in **Fig. 2(g)** exists in the inner domain. Therefore, the importance of the local curvature of the c-JJ is naturally assumed. This trend, however, should be checked with much higher statistics beyond focusing simply on three JVs.

We have investigated vortices with high statistics in an 800 x 800 nm$^2$ field of view. By considering distinct bound state energies for vortices, we define *Int*(+1.5 mV), integrated conductance between 0 mV and 1.5 mV [**Fig. 3(a-b)**]. While the obtained mapping [**Fig. 3(a,b)**] visualizes both AVs and JVs, the AVs alone can be further visualized based on mapping $g$(0 mV)/$g$(+5 mV) (**Fig. 3(c-d)**). Therefore, by combining $Int$(1.5 mV) and $g$(0 mV)/$g$(+5 mV), the positions of JVs along c-JJ are successfully distinguished. The number of obtained JVs is more than 80, including data under $B$ = 0.2 and 0.5 T (see [18] for the procedure details and also references [19,20,21,22,23,24] therein). Here, the red lines in **Fig. 3(a-d)** are along the normal direction to the domain boundary (black lines), which are also adjusted to cross the center of JVs. Systematic analysis is the next step, investigating spectroscopic evolution along red lines for all JVs.

To explain our quantitative analysis, one of the isolated JVs is focused [**Fig. 3(e)**]. First, the intensity of $Int$(1.5 mV) inside the curvature (left upper side) and outside the curvature (right lower side) along the red line are integrated within 25 nm from the junction. As in **Fig. 3(f)**, the integrated values are defined separately as $S_-$ (shaded by blue) and $S_+$ (shaded by red), respectively. Afterward, the asymmetric factor $\Delta S = S_+ - S_-$ is set to correlate with local curvature $1/R$ (see [18] for further details on the analysis). Uneven distribution of $\Delta S$ between positive and negative values means curvature plays a role in locating JV, and $\Delta S > 0$ is expected if a JV appears outside the junction. We have confirmed that the probability of $\Delta S>0$ is reaching ~ 75 % (out of more than 80 JVs) for 0.2 T and 0.5 T [**Fig. 3(g-h)**]. Furthermore, as in **Fig. 3(i)**, the trend of increasing $\Delta S$ (with keeping positive) with increasing local curvature ($1/R$) is seen. Since straight JJ ($1/R \to 0$) naturally results in JV positions symmetric against junction ($\Delta S = 0$), the linear fit of these data points (with crossing origin) is shown by the dotted line [**Fig. 3(i)**]. Regarding the existence of AVs and JVs near each focused JV, the realistic presence of intervortex interactions presumably distributes the data points, allowing for $\Delta S<0$ in **Fig. 3(i)**. Nevertheless, probability ~ 75 % for $\Delta S>0$ provides support to conclude that the curvature of the JJ plays a significant role in controlling the position of the JVs.

To understand the role of JJ curvature in positioning JVs, numerical simulations were performed by solving the Bogoliubov-de Gennes equation and gap equation self-consistently, as in the previous study [14]. In our simulation, totally five magnetic fluxes are positioned (see blue dots in **Fig. 4a**): at four corners and an additional one at the center of the field of view. Importantly, the curved JJ with controlled curvature ($1/R$) is set to cross the center position to understand the nature of JV on c-JJ (see dotted line in **Fig. 4(a)**). By defining $t$ (and $t'$) as the quasiparticle hopping energy within a domain (and domain boundary), the relative $t'/t$ is also adjusted to capture the experimental situation. From systematic simulations of $\Delta S$ with varying $1/R$ and $t'/t$ [**Fig. 4(b)**], reasonable $t'/t$ for modeling the real system is clarified to be around $t'/t \sim 0.05$ (see [18] for the details of systematic evaluation). Regarding the selection of $R$, while a larger curvature of $1/R$ is efficient for seeing the role of curvature, its value should be within the realistic number. As indicated by the vertical dashed line in **Fig. 3(i)**, focusing on $1/R = 0.05$ (nm$^{-1}$) with $R = 20$ nm meets this balance. By selecting $t'/t = 0.05$ and $R = 20$ nm [see arrow in **Fig. 4(b)**], the simulated LDOS(0 mV)/LDOS(+5 mV) visualizes AVs at four corners through ZBC anomaly [**Fig. 4(c)**]. A notable point is that this ZBC anomaly is almost absent at the center (but on c-JJ). Instead, the existence of the bound state energy around the gap edge is supported by simulation [arrow in **Fig. 4(d)**]. This theoretical simulation also confirms the validity of the experimental way to visualize JVs by focusing on gap edge energy [**Fig. 2** and **3**].

By mapping normalized LDOS, focusing on bound stat energy, JV is visualized to show predominantly outside of c-JJ [**(Fig. 4(e)**]. To understand this deformation, the key signature exists in the spatial variation of the phase of superconducting order parameters $\theta$ [arrows in **Fig. 4(f)**]. As seen in background color mapping, the spatial variation of local phase change $\Delta\theta(r) = \sqrt{(\frac{\partial\theta}{\partial x})^2 + (\frac{\partial\theta}{\partial y})^2}$ is more enhanced outside than inside of c-JJ [25]. Consequently, the selective formation of JV outside of c-JJ is realized, as opposed to inside. It should be stressed that the simulated JV represents $\Delta S > 0$, which reproduces major experimental observations out of more than 80 JVs [**Fig. 2 and 3**]. Therefore, we conclude that the curvature of JJ has an intrinsic impact on the control position of JV.

Finally, mentioning the small *t'*/*t* (= 0.05) employed in our simulation is intriguing. It should be stressed that this value is even smaller than the previous study on disconnected atomic sheets [14]. In $LiTi_2O_4$, it is worthwhile to point out that distinct electronic symmetry between adjacent domains can be a potential source of anomalous suppression of hopping in our system [16,17]. While this study employs the simplest s-wave pairing symmetry for simplicity, the possible unconventional pairing symmetry in LTO can be effectively captured through an anomalously reduced hopping parameter *t'*/*t*. We want to clarify this for future studies.

In summary, this is the first report to elucidate the critical role of curvature on the Josephson junction to control the position of the Josephson vortex. Given that our simulations utilize conventional parameters, the insights provided by our study can be qualitatively applied to a wide range of superconductors.

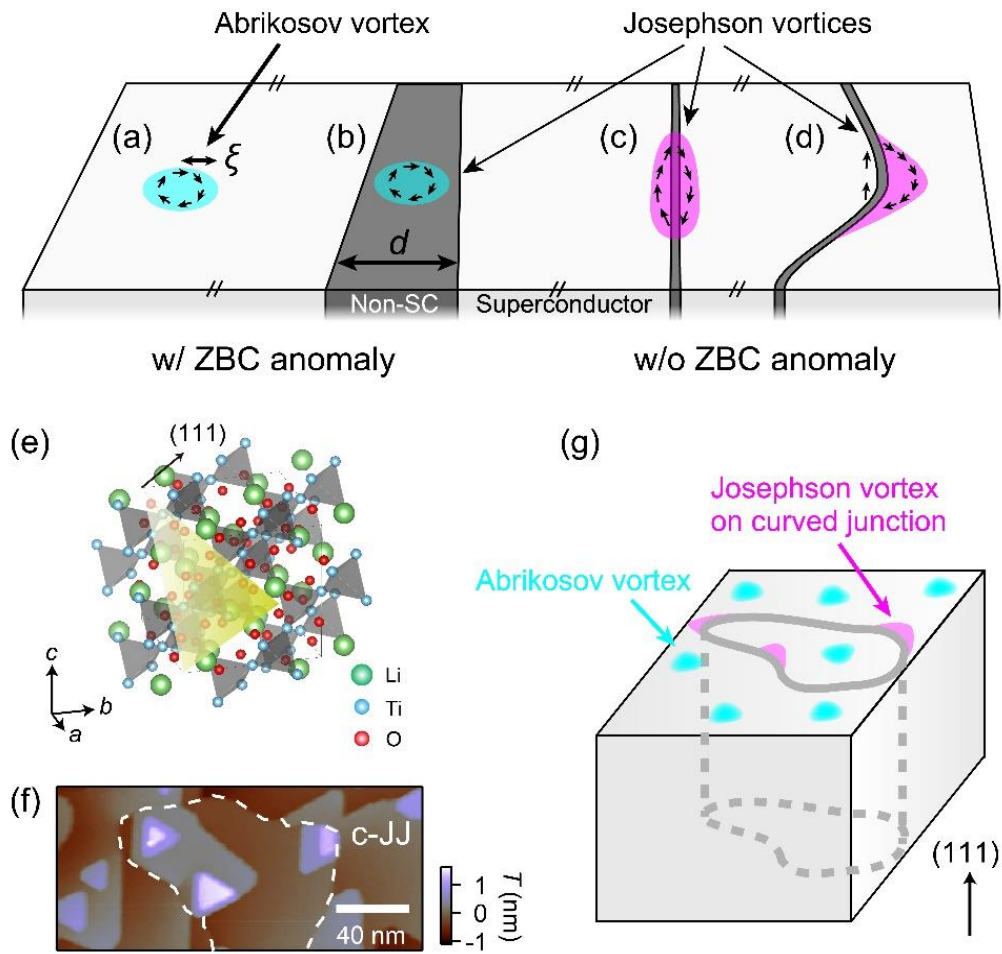

**Fig. 1.**

(a)–(d) Schematic representation of superconducting vortices. See the main body for details. White (black) regions represent superconducting (normal/insulating) regions. The black arrows represent net magnetic flux around the core (phase of the order parameter). (e) Crystal structure of $LiT_2O_4$ drawn by VESTA [26]. (f) A representative STM topograph showing a domain structure (see the while curve). The feedback condition is 7 mV/200 pA. The curved Josephson junction is represented by c-JJ. (g) Schematic representation of the Abrikosov and Josephson vortex investigated in this study. The blue triangles represent the Abrikosov vortices, while the pink contrast represents the Josephson vortex formed near the crystalline domain boundary (shown by the gray curve).

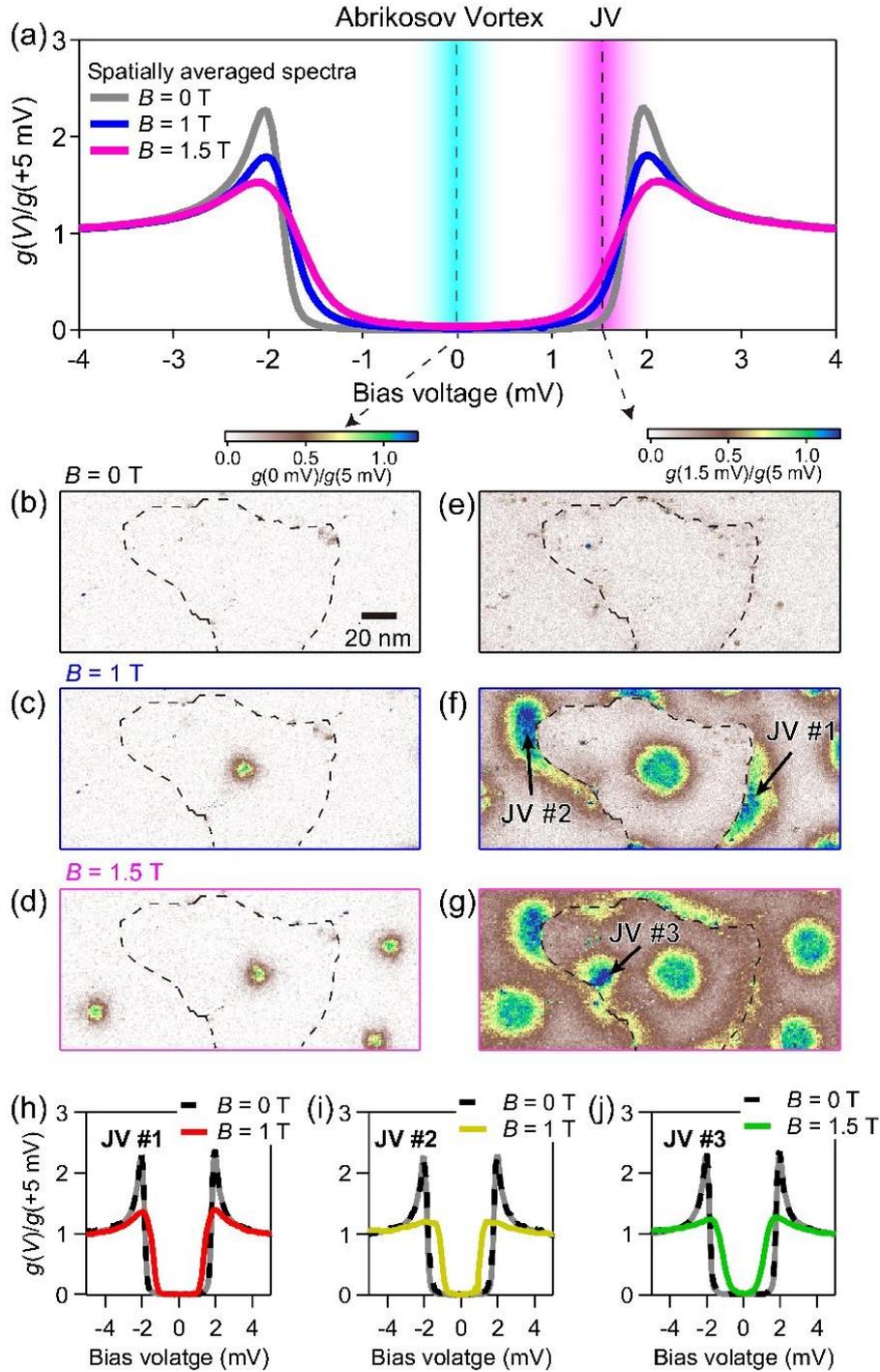

**Fig. 2.**

(a) Spatially averaged $dI/dV$ spectra taken at $B$ = 0,1, 1.5 T. Vertical dashed lines represent the energy where we map out the local density of state contrast for each field. (b-d) $g$(0 mV)/$g$(5 mV) maps taken with $B$ = 0 (b), 1 (c), and 1.5 T (d). (e-g) $g$(1.5 mV)/$g$(5 mV) maps taken with $B$ = 0 (b), 1 (c), and 1.5 T (d). The black dashed lines indicate the c-JJ. (h-j) $g$(V)/$g$(5 mV) spectra taken on JV#1 (h), #2 (i), and #3 (j) as indicated in f and g (colored spectra). Dashed black (gray) spectra represent the spectra at $B$ = 0 T, taken on the same locations as the JV#1-3 observed (spatially averaged spectra at $B$ = 0 T).

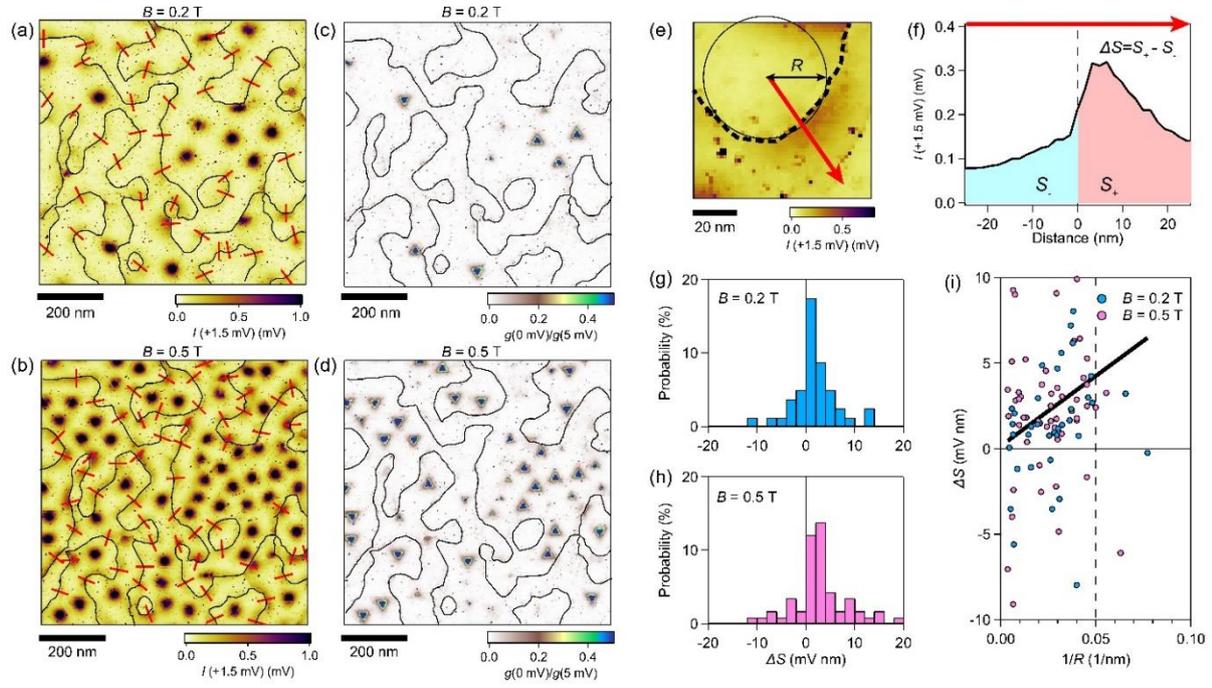

**Fig. 3.**

(a,b) *Int*(1.5 mV) maps at 0.2 and 0.5 T, respectively. The black lines indicate the crystalline domain boundary. Red lines indicate the location of the JVs. (c,d) $g(0\ \text{mV})/g(5\ \text{mV})$ map at 0.2 and 0.5 T, respectively. (e) A representative JV appeared in *Int*(1.5 mV) map at 0.5 T. The thick curve indicates the domain boundary. The line profile is taken along the red line to examine the position of the c-JJs quantitatively. The local curvature radius ($R$) is estimated to be 25 nm. (f) The line profile along the red line in e. The horizontal origin is set to the location of the domain boundary. (g,h) The histograms of $\Delta S$ at 0.2 and 0.5 T. (i) Relation between the inverse of local curvature radius ($1/R$) and $\Delta S$. The dashed black line is the guide to the eye, showing a correlation between $\Delta S$ and $1/R$.

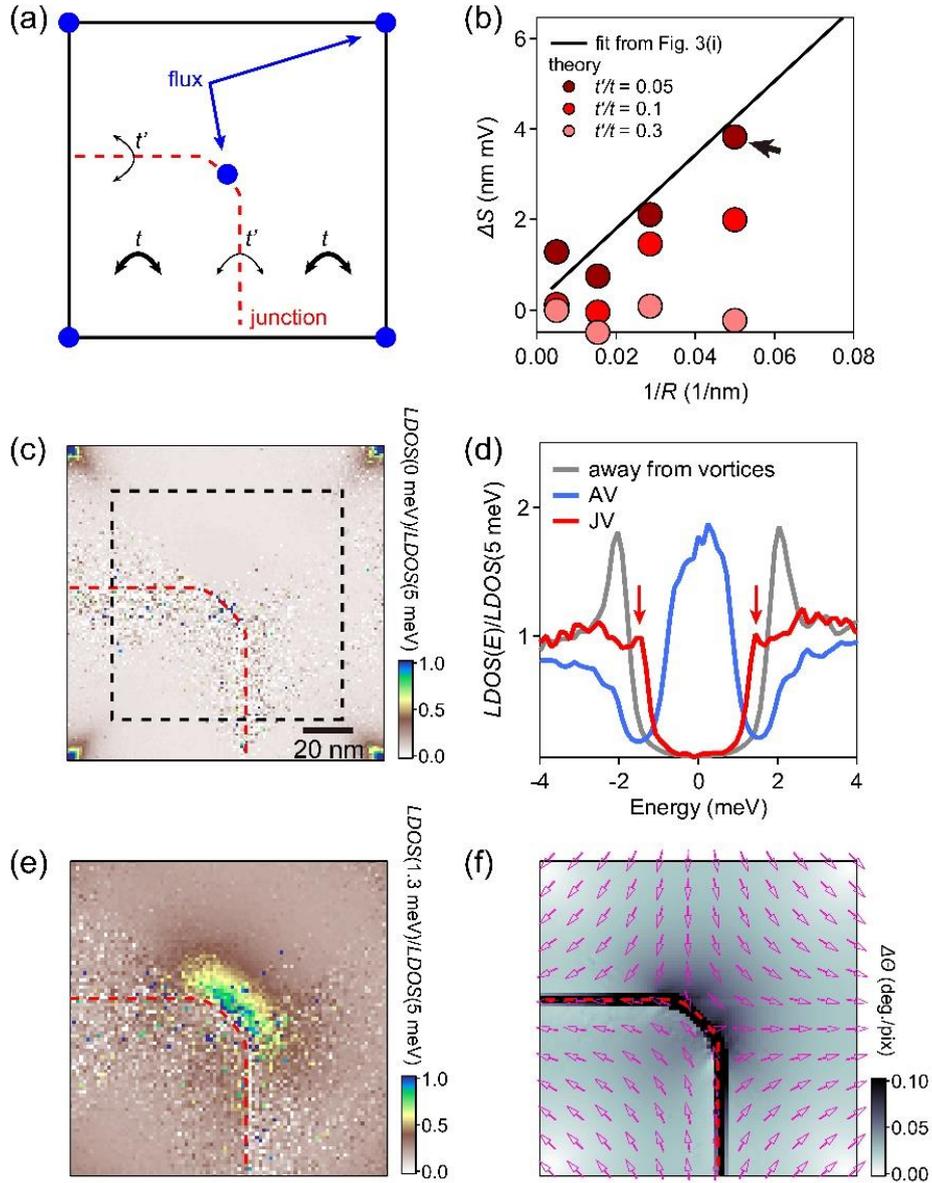

**Fig. 4.**

(a) Schematic representation of the simulation. The blue dots represent initial position of magnetic flux. The red dotted line indicates curved Josephson junction. The quasiparticle hopping within the domain (t) and crossing junction (t') are also represented. (b) The relation between $\Delta S$ and $1/R$ with varying $t'/t$. The linear fitting obtained from experiments [Fig. 3(i)] is also shown as the black line. (c) $LDOS$(0 meV)/$LDOS$(5 meV) map. (d) $LDOS$ spectra taken on the JV (red) and on the AV (blue), and far away from vortices (gray) in (c). (e) $LDOS$(1.3 meV)/$LDOS$(5 meV) maps. The red dashed lines indicate the JJ. (f) Spatial evolution of the phase of the superconducting order parameter in the curved junction ($R$=20 nm). The pink arrows represent the phase (from 0 to $2\pi$), whereas the color image behind represents the differentiation of the phase ($\Delta\Theta$). See the main text for more details. The area of (e) and (f) are indicated by the dashed square in (c).

[25] One may suspect there are multiple solutions in the phase map to reproduce the LDOS contrast. However, we would emphasize that the presented phase map satisfies two important conditions: it reproduces a similar contrast in the LDOS map to the experimental one, and it minimizes the total energy. Thus, the phase map as shown in **Fig. 4(f)** is one of the reasonable solutions.

[26] K. Momma, and F. Izumi, J. Appl. Crystallogr. **44**, 1272-1276 (2011).